\date{\today}
\documentclass[12pt]{article}
\usepackage{url}
\usepackage{graphicx}
\usepackage{float}
\usepackage{sansmath}
\usepackage{xcolor}
\usepackage[affil-it]{authblk}
\begin{document} \sloppy
\renewcommand\Authfont{\fontsize{12}{10}\selectfont}
\renewcommand\Affilfont{\fontsize{6}{0}\itshape}
\title{\huge The Scenario Culture}
\author[1]{Edward Wheatcroft}
\author[1]{Henry Wynn}
\author[2]{Chris J. Dent}
\author[3]{Jim Q. Smith}
\author[4]{Claire L. Copeland}
\author[5]{Daniel Ralph}
\author[6]{Stan Zachary}
\affil[1]{\tiny London School of Economics and Political Science}
\affil[2]{\tiny University of Edinburgh}
\affil[3]{\tiny University of Warwick}
\affil[4]{\tiny University of Sussex}
\affil[5]{\tiny University of Cambridge}
\affil[6]{\tiny Heriot-Watt University, Edinburgh}
\maketitle
\noindent
\normalsize
\begin{abstract}
\noindent
Scenario Analysis is a risk assessment tool that aims to evaluate the impact of a small number of distinct plausible future scenarios. In this paper, we provide an overview of important aspects of Scenario Analysis including when it is appropriate, the design of scenarios, uncertainty and encouraging creativity.  Each of these issues is discussed in the context of climate, energy and legal scenarios.
\end{abstract}

\section{Introduction}
Scenario analysis (SA) is a technique that assesses the impact of a small set of distinct scenarios, with the aim of gaining insight into how the future might evolve under different defined conditions.  The rationale of SA differs from that of forecasting in that scenarios are not extrapolations of the past but are built on certain plausible underlying assumptions about the future.  Stress testing, a term which is often used interchangeably with SA, is a similar concept.  However, the aim is to consider extreme scenarios in order to assess the effects of unlikely but high impact events.   \par

The aim of this paper is discuss various aspects of SA and to relate them to three areas of application. We do not claim to offer a coherent theory of scenarios, as, to some extent, the field is still developing.  However, we hope that this paper will go some way towards its development.  The three areas of application are (A) climate, (B) energy and (C) litigation in insurance, and these will be referenced where appropriate.  For (A), we focus on the Intergovernmental Panel on Climate Change's Representative Concentration Pathways \cite{van2011representative}, which assume different concentrations of greenhouse gases in the earth's atmosphere.  For (B), we use on the UK National Grid's Future Energy Scenarios \cite{Future_energy_scenarios} which consider different energy futures for the UK both in terms of decarbonisation of the economy and decentralisation of energy production.  For (C), we briefly define our own scenarios, based on our experience from a number of projects in criminal and civil law, including a project related to accident insurance.  Details of each of these are given in the appendix. \par

In the case of energy, in addition to the National Grid scenarios described above, we draw on our own experience on two European projects in the area of district heating \cite{Celsius,REUSEHEAT} and from individual research projects consolidated at the Alan Turing Institute, particularly the project Managing Uncertainty in Government Modelling (MUGM) \cite{MUGM}. For the litigation case, we exploit our experience working with a number of parties interested in dealing with uncertainty surrounding the development and uptake of driverless cars. \par

Purposes of SA can broadly be placed into two different categories.  In the first, often referred to as \emph{Scenario Planning}, the aim is to assess which future events are plausible and what knock-on effects they might have and to encourage planners to think about the future and what risks and opportunities they capture.  This way of thinking is common in disaster planning, for example.  The second use of scenarios is in \emph{decision making} in which scenarios are used to aid a specific decision. For example, in energy planning, the impact of different levels of future energy demand might be considered. \par

SA is considered by many organisations to be an important tool in the evaluation of risk.  The UK Government's Green Book, which provides guidance on how to appraise and evaluate policies and projects, recommends SA for `significant future uncertainties' caused by potential technical, economic and political issues \cite{treasury2018green}.  The UK Government's Aqua Book, which provides guidance for producing analysis for government, recommends that analytical models should be tested `under a variety of scenarios' \cite{treasury2015aqua}. The European Union has also made use of scenarios in its energy planning.  The 2011 EU Energy Roadmap defined a set of scenarios targeting the reduction of greenhouse gas emissions by 80 percent or more by 2050 \cite{european2012energy}. Other examples of organisations that recommend or use SA are the European Chemicals Agency \cite{ECHA} and the European Food Safety authority \cite{benton2019using}. \par

Scenario planning has been used in a wide variety of different fields and has a considerable history.  In 1973, Peter deLeon of the Rand Corporation wrote a report detailing his experiences of using scenarios in `games' centred around political and military situations \cite{deleon1975scenario}.  The games were designed with two `teams' who were faced against each other and a game-manager in charge of setting various scenarios for the teams.  The main aim of the games was to put people in unpredictable positions and therefore to encourage them to practice their decision making in a challenging environment. \par

In climate science, the level of future greenhouse gas emissions is a major uncertainty that depends on a range of political and societal aspects.  Given the unpredictable nature of human behaviour, SA is a useful tool in this context because it allows the impacts of climate change to be assessed in multiple settings. Rather than attempting to predict future behaviour, the approach is simply to assess the impact were the assumptions under each scenario to emerge. \par

In fact, scenarios have been used in climate science for a number of decades.  A 1988 study assessed the impact of three emissions scenarios on global and local temperatures \cite{hansen1988global}. The Intergovernmental Panel on Climate Change (IPCC) first produced its IS92 scenarios in 1992 and these were presented in the second IPCC assessment report. IS92 consists of six scenarios each featuring differing assumptions regarding economic, social and environmental conditions \cite{houghton1995climate}. Later, in 2000, the IPCC's `Special Report on Emissions Scenarios' (SRES) was produced and designed to improve on a number of aspects of the IS92 scenarios such as the assumed nature of global economic restructuring and the rate of technological change \cite{nakicenovic2000special}. The SRES were superseded in 2014 by Representative Concentration Pathways (RCP) \cite{van2011representative} which are defined on the basis of different radiative forcings, that is the effect of increased emissions on the make up of the atmosphere, rather than actual greenhouse gas emissions.  The RCP scenarios are described in the appendix. \par

SA is rather less common in the legal context.  However, the Hague Institute for Innovation of Law (HiiL), a social enterprise that aims to increase the accessibility of the justice system, has produced a report entitled `Law Scenarios to 2030' which explores possible scenarios for the global legal industry. In that report, the four scenarios are differentiated on the basis of different levels of internationalisation (i.e. to what extent laws are made internationally) and whether laws and regulations are predominantly decided at a state or private level.  It is acknowledge in the report that scenarios are `relatively unknown as a working tool for lawyers and jurists'.  However, the value of scenarios as a tool to `allow for uncertainty and account for it' is argued, underpinning the potential value of SA in this area. \par

SA is widely used in the planning and construction of infrastructure projects.  For example, downscaled climate projections have been used as scenarios when considering the impact of flooding \cite{jenkins2017assessing} and future demand for air conditioning \cite{Future_of_cooling}.  Another well known set of scenarios are defined by Shell’s Global Business Environment team which produces different scenarios of future energy supply and demand using their Global Supply Model \cite{Shell}.  The Shell scenarios differ in their underlying assumptions regarding hard-to-forecast factors such as geopolitical events and improvements in technology.  As discussed further in the appendix, the UK National Grid also annually produces future energy scenarios on behalf of the UK Government \cite{Future_energy_scenarios}.  Details of the scenarios are disseminated to the public and are available online \cite{NG_scenarios}.  The EU Reference Scenario produces its own energy scenarios which are typically used in decision making for EU-wide initiatives \cite{EUreference}.  SA has also been used in the context of District Heating and Cooling (DHC).  For example, in \cite{Chalmers}, four scenarios were defined to assess the effects of greater energy price volatility and improvements in thermal storage on the future prospects of a district heating scheme in Bor\aa s, Sweden. \par

\section{Good practice}
The following sections each relate to different aspects of SA and we discuss what we believe to be good practice in each case. The aim is then to provide some views on applying SA in practical decision making.  Each one is discussed separately with reference to the climate, energy and legal scenarios defined in the appendix and in a wider context.  \par

\subsection{Designing scenarios}
An important aspect of SA concerns the underlying basis on which scenarios are designed. A number of approaches are taken, in practice:

\begin{enumerate}
\item Advantageousness: scenarios are defined in terms of how advantageous or unfavourable each one would be if it came to fruition.  For example, scenarios labelled \emph{optimistic}, \emph{pessimistic} and \emph{neutral} might be defined.  This has been done in the context of assessing the future prevalence of smoking \cite{basu2011projected}, for example. 

\item Event based: scenarios are defined based on the occurrence or nature of some underlying event.  For example, at the time of writing in which Britain's future relationship with the European Union has not been decided following the 2016 referendum, different scenarios might be defined to assess the impact of this uncertainty.  Scenarios in this case might be defined on the basis of different policy directions such as Britain staying in the European single market, the negotiation of a new customs union between the UK and the EEA or a second referendum being held and Britain opting to remain. 

\item Level based: scenarios are defined on the basis of different levels of some quantitative variable.  For example, the IPCC's Representative Concentration Pathways (RCPs) are defined on the basis of different levels of greenhouse gases in the atmosphere over time. In energy planning, scenarios might be based on different future electricity prices and, in the context of motor insurance, legal scenarios might be based on the proportion of self-driving cars on the road over time. \par
\end{enumerate}

Note that, whilst each of the above provides a basis for scenario design, it is common for a combination of these to feature simultaneously within the scenario itself.  For example, consider a set of scenarios designed to assess the viability of a district heating scheme in which one of the scenarios is that government with a significant green agenda achieves power.  Whilst such a scenario might be considered an `event based' scenario, the assumptions that make up that scenario might be a combination of level based aspects such as carbon taxes and event based aspects such as total bans on certain technologies or fuels. \par

An important question in SA regards the number of distinct scenarios to be considered.  In order to avoid overcomplication and confusion, a maximum of three to five is often recommended \cite{aaker2010strategic}.  

In \cite{fsb}, a number of conditions for effective scenario design are defined.  Whilst these are discussed in the context of climate related risk, we believe they are useful for the design of scenarios in other contexts.  Each of the conditions are defined and described below:
\begin{enumerate}
\item Plausible - a scenario should be plausible and come with a narrative justifying each event or change in the underlying assumptions.
\item Distinctive - the different scenarios should be distinctive enough in terms of the key factors for there to be a clear difference between them.
\item Consistent - Interaction between key factors should be taken into account.  For example, macroeconomic factors may impact important aspects of the scenarios simultaneously.
\item Relevant - each scenario should be relevant in terms of giving a specific insight into the future (e.g. the government increases spending on green projects and subsidies).
\item Challenging - scenarios should challenge the conventional view on things that may affect the project in question.
\end{enumerate}
Consistency is particularly important and, arguably, SA is well placed to deal with associations between variables. In sensitivity analysis, defining such associations can be a difficult task since they can be complex.  Since SA only considers a small number of potential pathways, it is much easier to account for such dependencies. \par

Consider the following case from the energy sector.  Suppose that a local authority is considering investing in a district heating scheme but there is an upcoming election in which a green-minded government could plausibly win power.  Whilst it is often difficult to predict government policy, there are strong rumours that a government formed by that party plans to increase taxation on greenhouse gas emissions and increase subsidies on renewable energy.  These are two effects caused by one underlying factor, i.e. the election of a green-minded government. It is therefore important that these dependencies are accounted for in the design of the scenarios.  \par

An important question in scenario design concerns whether scenarios should be designed with a specific decision in mind.  This depends heavily on the application.  The IPCC's Representative Concentration Pathway scenarios, for example, are designed to assess the impact of different levels of greenhouse gas emissions.  As such, the decision to be made is to what extent emissions should be reduced over this period. The National Grid's future energy scenarios are built for use by the UK Government in its energy planning.  In both cases, however, the scenarios are in the public domain and available for use by anyone that wishes to use them.  In the case of motor insurance, SA is less widely used and thus general scenarios for issues like the onset of driverless vehicles are not widely available.  Therefore if an insurance company wishes to apply SA, they will need to define the scenarios for the specific question(s) they wish to answer.  In some sense, insurance policies are designed specifically to take into account scenarios and their risk. For example, there is current interest as to whether climate change will affect flood insurance. \par  

The decision as to whether scenarios should be designed with a specific question in mind often ultimately comes down to their intended purpose. Generalised scenarios are often more efficient in terms of costs since they can be reused for a range of purposes.  However, designing scenarios for general use often comes at a cost in that they are less relevant to the question at hand. \par

\subsection{Control}
The design of scenarios is influenced by the level of \emph{control} that the decision maker has.  This concept is familiar from risk management where the ability to control a risk and to mitigate the effects of an event are critical \cite{kliem2019reducing}. Consider, for example, a municipality in the United Kingdom that wishes to decide whether to proceed with a proposal for a district heating scheme.  To help understand the risks associated with the decision, they decide to utilise the National Grid's energy scenarios and assess their impact on the viability of the project.  Some potential elements such as government policy are out of the control of the city whilst, for others, they have direct control.  For example, changes in the wholesale electricity price cannot be controlled by the city and thus can only be mitigated, although ideally they should be incorporated into mathematical modelling at the local level.  Subsidies for solar panels, on the other hand, might be in the direct control of the city.  This distinction brings up a subtle difference in potential policy actions.  In the former case, the city can only mitigate but, in the latter, the city is able to directly impact the nature of key effects to the project of interest. \par

The level of control over the scenarios differs over our three examples.  For the case of the IPCC's Representative Concentration Pathways, the level of control clearly differs depending on who is making the decision.  For example, it could be argued that world governments as a whole have a significant level of control over greenhouse gas emissions, national governments have some control and businesses have limited or no control. For the UK National Grid's Future Energy Scenarios, the UK government has a great deal of control in that they are able to design policy whilst small businesses have little control.  Finally, for scenarios regarding driverless vehicles (appendix C), the government has a high level of control should it wish to exercise it (depending on its attitude towards regulation) whilst insurance companies may have limited control. \par  

\subsection{Long term versus short term objectives}
SA differs significantly according to the timescale of the decision to be made or action to be taken.  Similarly to forecasting, the longer the time horizon of interest, the more uncertainty.  In the design of long term scenarios, such as in climate science in which scenarios can reach fifty years or more into the future, it can be extremely difficult to identify plausible pathways.  It is much easier for short term objectives, on the other hand, since one can more readily imagine how events might develop.  This is particularly true when the same decision is made repeatedly. Consider, for example, a case in which the objective is to ensure that a country is able to meet energy demand over the following two-year period.  This is a decision that is likely to have been made regularly but is still subject to qualitative factors such as political changes or geophysical risks. \par

\subsection{Scenarios and expert judgement}
Scenario design is a challenging process that can involve the identification of unprecedented future events.  This often requires a great deal of creativity and insight that may only be achievable with the input of experts in the field.  The benefit of asking experts to help design scenarios is twofold.  First, asking an independent party to design a set of scenarios separates scenario design from scenario planning.  This discourages the tendency to design scenarios that planners are already familiar with and are confident that they can handle.  Second, an expert may be better equipped to identify a wide range of different possible scenarios and, as a result, planners and decision makers should be in a better position to deal with future events. \par

The observation that important aspects of scenarios in Scenario Planning can be neglected is one that has been made by other authors.  Referring to this issue, according to \cite{meissner2017quantifiying}, 
\\
\\
\small `This can be critical for organisations if it leads to an inertia in internal judgement, resulting in blind spots or a failure to see weak signals in the firm’s periphery.'
\\
\\
\normalsize
In that paper, a methodology is described with which to integrate external expert judgement into Scenario Planning. \par

Expert judgement can also play a role in estimating probabilities or ranking the likelihood of different scenarios.  This is particularly true in estimating probabilities of qualitative events that are hard to model.  Expert judgement, in this setting, typically involves the estimation of a probability distribution by one or more experts in a field.  When multiple experts are involved, the question of how to incorporate all of this information into a decision or probability estimate is an important one.  The Sheffield Elicitation Framework (SHELF) \cite{gosling2018shelf} provides extensive guidance on elicitation of probabilities from multiple experts.  Under the framework, multiple experts are asked to form independent probability estimates.  These then forms the basis of discussions between the experts, after which an impartial observer forms a `consensus' distribution. \par 

A concern during expert elicitation is that the experts themselves may be poorly calibrated, that is, despite their expertise, they are typically unable to form accurate probabilities.  Roger Cooke's classical model \cite{cooke1988calibration} asks experts to provide judgement on past events in which the outcome is not revealed \cite{colson2018expert}.  This allows for the ability of each expert to be assessed and potentially recalibrated to remove biases.  In theory, this approach could be extended to assess the probabilities of each scenario in SA, though some qualitative approach may also be needed. \par

In disaster planning, the job of building scenarios is often given to a set of experts who then present those scenarios to decision makers who are asked to assess the effects and, crucially, suggest potential mitigating action. This is commonly done in the planning for a potential nuclear accident, for example.  The State-of-the-Art Reactor Consequence Analyses (SOARCA) project has been set up by the U.S. Nuclear Regulatory Commission to provide realistic scenarios regarding potential nuclear reactor accidents \cite{SOARCA}.  Users are able to define scenarios based on their likelihood of occurrence whilst computer modelling is used to predict how a reactor might behave under such conditions \cite{SOARCA_report}. \par

Many examples of expert judgement in SA can be found in the scientific literature.  For example, subjective probabilities were elicited from multiple experts regarding the risk to participants in the annual Pamplona bull run in Spain \cite{mallor2008expert}. Expert Judgement was also used in \cite{heiko2010scenarios} in the form of the Delphi method to combine the judgement of 30 CEOs of service providers to design scenarios for the logistics industry up to 2025. \par

\subsection{Creativity and scenarios}
Effective risk assessment requires creativity in order to identify potential future risks, particularly if the timescales are long.  This is particularly important in disaster risk management where risks are constantly emerging and, if ignored, can be highly damaging. The importance of creativity is summed up by the title of Herman Kahn's book `Thinking About The Thinkable' which, in his case, largely focuses on scenarios of potential nuclear conflict \cite{kahn1962thinking}. \par

A variety of different techniques can be used to encourage decision makers to be creative when identifying potential risks. One way to encourage creativity is to try and \emph{immerse} people in a situation as much as possible so that they can more readily imagine potential risks or opportunities. The use of games by the Rand Corporation in the 1970s was designed to give participants a stake in the decision to incentivise them to think carefully as to how the situation might develop \cite{deleon1975scenario}. \par

The onset of modern technology has enabled far more realistic depictions of reality.  \emph{Virtual Scenario Planning} \cite{mcwhorter2014initial}, in which participants are placed into a virtual reality world and asked to make decisions in real time, is growing quickly in popularity.  One of the main advantages of Virtual Scenario Planning is that it enables participants to visualise different scenarios, creating a far more immersive experience than one in which the situation is simply described.  Potential use of virtual reality in the planning of smart cities is reviewed in \cite{jamei2017investigating}. \par

The use of scenarios to enhance creativity has strong analogies with the use of conceptual design in areas such as architecture and product development.

\section{Uncertainty}
A key question in SA is whether scenarios should come with a probabilistic estimate attached.  The plausibility of doing this depends on a number of factors.  One problem is that is often difficult to design scenarios that span the probability space, that is provide an exhaustive set of possible outcomes.  Probabilities in this case are ill-defined. \par 

Consider, for example, a set of energy scenarios in which the electricity price is a factor.  Typically, in each scenario, a price would be defined as a function of time.  In reality, it is highly unlikely that the electricity price would develop exactly as described in any of the scenarios and therefore placing a probability on any of them would be a difficult and, arguably, meaningless task.  One way in which to attempt to avoid this situation is to define scenarios using different defined levels of some variable.  For example, energy scenarios could be defined with a ``high'', ``medium'' or ``low'' electricity price.  This causes further complexities, however, as the question of how these translate into model parameters is hard to answer.  

Even with a set of scenarios that span the probability space, assessing the probability of each can be extremely difficult, particularly when long time scales are considered.  Dependencies between scenario elements can quickly make the calculations complex and probabilities inaccurate.  In many cases, the value of placing a probability on a scenario is questionable. \par

In some cases, the decision maker has some control over how a scenario develops.  This can make things even more complex since the probabilities of each scenario are impacted by the actions taken by the decision maker.  Whilst this effect complicates matters, it should be noted that real option theory aims to answer the question of how to deal with such situations \cite{trigeorgis2017real}. \par

Given the difficulty of assigning probabilities to scenarios, it is interesting to consider how the results should be presented in a decision making context.  Inevitably, a decision maker will want to gain some insight into the relative likelihood of each scenario.  If this is not possible, analysis can be performed under each scenario and the results presented for each one separately.  Discussion of the relative likelihood can then take place.  In some cases, the decision maker themselves may be informed enough to have their own insight on the matter.  One simple way of presenting the relative likelihood of each scenario is to rank them in order of their plausibility. \par

The IPCC's Representative Concentration Pathways do not come with probability statements attached.  This is likely because of the difficulty of predicting the political and social behaviour of humans over a long period of time.  However, there is also another explanation in that, since the IPCC is an organisation set up between national governments, the likelihood of each scenario is dependent on the impact of the scenarios themselves. \par

The UK National Grid does not provide probability statements regarding scenarios.  In 2014, its outlook manager Gary Dolphin stated \cite{carbonbrief} that 
\\
\\
\small `We do not apply probabilities to any of our scenarios. We like to believe each scenario is equally plausible and that they collectively define an envelope within which the true future lies.'
\\
\\
\normalsize
One could argue, however, that stating that each scenario is equally plausible endorses the view that none should be treated as any more likely than any other. \par

In our view, the mutually enhancing relationship between uncertainty and SA is likely to be fruitful.  Although uncertainty remains an opaque and ill-defined subject, it is commonly agreed that there is a gap to be filled between formal probabilistic methods and non-probabilistic methods such as those used in areas in which there is `radical uncertainty' \cite{CRUISE}.  It seems to us that SA provides a useful contribution to closing this gap.  A simple, but possibly attractive, way of thinking about these matters is that there are two broad categories of uncertainty: one containing SA, global sensitivity and long term forecasting and the other a much better developed area of local sensitivity and short term decision making. \par

\section{Scenarios and model error}
Scenarios are typically used to test the effects of different model input assumptions on model output. It is now generally agreed that no mathematical model can perfectly represent the behaviour of the world and therefore can never be expected to provide perfect predictions of future events.  In SA, there would therefore be an inevitable discrepancy between model output and the real world, even if the assumptions developed exactly as defined under the scenario.  This discrepancy can be particularly large when unusual scenarios are considered.  In such cases, the model may be ill-suited to providing realistic projections.  SA can, in fact, help to uncover weaknesses of models.  If a set of scenario assumptions in a model provide clearly unintuitive results, this may provoke a closer look at the model to attempt to understand the cause. \par

Consider, for example, a case in which SA is used to assess the effects of a long electricity outage. If the aim is to assess the effects of such an outage occurring on a weekday but the model is built on the basis that it occurs on a weekend, given that society's behaviour is typically different on weekends and weekdays, the model is unlikely to be fit for purpose.  The discrepancy between the model output and what might be expected to happen intuitively may, however, be able to uncover this model inadequacy. \par 

Ideally, a mathematical model of a physical, social, economic or environmental activity would incorporate the facility to model all scenarios thought likely.  This is probably economically infeasible but it is recommended that models are built in modular form so that, as new scenarios come into view, the model can be extended. Having a legacy model which cannot deal with an important scenario is inflexible. Insurance companies needing estimates of increasing precipitation due to climate change leading to runoff from mountains will not be impressed if the mountains are not included in the climate models. \par

\section{Scenarios in decision making}
Perhaps the most important issue in SA is how scenarios can be used to make decisions.  If the scenarios span the probability space and accurate probabilities of each scenario can be calculated, the expected cost or utility can be found using standard methods. A decision can then be made that optimises the variable.  In practice, as described above, it is often unreasonable, and sometimes even ill-defined, to assign probabilities to each outcome. In that case, this approach is not feasible. \par

Scenarios are sometimes defined as different pathways to achieve some defined goal.  This can then provide a range of options for policy making.  An example of scenarios defined in this way can be found in the Greater London Authority's 1.5C Compatible Climate Action Plan \cite{GLA}.  In that, four different scenarios are defined by which the goal of a carbon neutral London by 2050 can be expected to be achieved. \par

At the time of writing, plans are being drawn up by the Bank of England to require businesses to test the impact of different climate scenarios including a `business as usual' scenario in which greenhouse gas emissions are not significantly curtailed, and one in which the UK Government achieves its aim of transitioning to a carbon neutral state by 2050.  It is not clear at this stage what conditions will be imposed on businesses but there will likely be a direct impact on decision making.  The requirement to use sensible scenarios may be added to corporate governance rules. \par

The allocation of probabilities to scenarios places the theory within the area of multiple modelling, either Bayesian or non-Bayesian.  This is very close to metamodelling which grew out of a need to handle multiple models, for example run by different laboratories or research groups, on essentially the same entities.  The analogy is strong because, if all scenarios are housed within a single metamodel, then they can be represented by different levels of inputs.  Another strong analogy stems from saying that scenario analysis is a type of global sensitivity in which inputs, or even structural aspects of the model may be changed, as opposed to local sensitivity analysis where inputs are varied around some central or nominal value.  The analysis is close to a multi-level analysis with the indicator of the scenarios providing another level to which all the individual scenario models are attached.  \par

In a Bayesian context, the judgement of the probability of the occurrence of a particular scenario has the status of a prior probability.  Then, at least theoretically, the probabilities can be updated as the future develops and further data arise.  Indeed, should one or more scenarios seem more likely then one could alter data collection to concentrate on them. \par

In this regard, there is a flavour of play-the-winner-rules which have a long history in statistical decision theory. The theory which covers the Bayesian approach is usually called Bayesian Model Averaging (BMA).  This is a good label because it describes how the model behaves: the metamodel parameter estimates are essentially posterior averages of posterior model estimates.  Although coherent in a Bayesian fashion, it may not be appropriate to model the real world as an average of several `worlds'.  If the scenarios are mutually exclusive, then, at most, only one will occur and this will not then be an average.  In fact,if the scenarios are non-exhaustive, it may be that none of them occur. In such a case, we may put a prior probability on a `something else' scenario.  Classical non Bayesian meta-analysis is more like multiple hypothesis testing, in which quantities such as false error rates and similar are studied.  It is important in medicine where one of a number of different drugs may be efficacious.  This is a world in which one proceeds as if a particular hypothesis is true such as marketing the best drug for some ailment.  After all, in emergency situations, one has to do something and even doing nothing is doing something. \par

\subsection{Least Worst Regret}
In decision theory, \emph{regret} is the emotional reaction to information about the best course of action after having taken a decision. Least Worst Regret (LWR) is an approach in which decisions are made with the aim of minimising regret. An often argued benefit of LWR is that the probabilities of each scenario are not required.  Regret is defined as the difference in cost between the decision taken and the optimal decision. The difference is `regretful' since, with a decision, the cost could have been lower. In its Electricity Market Reform Report 2015 \cite{Nat_grid_report}, the National Grid echos the clam that probabilities are not required by stating that:
\\
\\
\small `One benefit of this approach is that it is independent of the probabilities of the various potential future outcomes and therefore it can be used when the probabilities of these outcomes are unknown, providing that the cases considered cover a range of credible outcomes.' 
\\
\\
\normalsize
LWR is built around the principal that humans tend to be risk averse and seek to avoid regret as much as possible \cite{bell1982regret}.  This idea is closely linked to \emph{Prospect Theory} in behavioural economics which describes how the negative utility from a perceived loss tends to be higher than the positive utility from the equivalent gain.  Regret is a powerful emotion with strong psychological effects that can have a big effect on a human.  The effect of regret can also be significant for an organisation where the risk of being responsible for a big `missed opportunity' can be high.  \par

It has been argued that the ability of LWR to aid useful decision making depends on a number of factors such as the cost function, the decision to be made and the probability of each scenario and thus should not be used uncritically and without careful consideration of these factors \cite{zachary2016least}. \par

In our opinion, the evidence justifying the use of LWR is weak.  The avoidance of discomfort to the decision makers that result from cognitive biases is not a sound basis for making a decision, nor is the seemingly deliberate attempt to avoid attaching probabilities to scenarios. Likewise, whilst, inevitably, reputational issues will almost certainly have an impact in decision making, it seems odd to embed the biases of the decision maker in the methodology behind the decision itself.  Arguably, it would be much better for the decision maker to focus on justifying their decision more scientifically. \par

Another criticism of LWR is that it can discourage the consideration of unlikely scenarios since there is nothing in LWR that requires differentiation between the likelihood of each one.  If a scenario is omitted because it is considered to be unlikely, some assumption about the likelihood of a scenario has effectively already been incorporated into the decision.  The question of how that information can be used better then inevitably arises. \par

\section{Discussion}
There is no doubt that the use of scenarios is gaining ground and that this can be attributed to a generalised anxiety concerning the future.  This is happening at a national and local level, and in many sectors from how climate change might affect flooding to sources of volatility in financial markets. Perhaps the most important question is how SA is actually to be used in or adapted for real decision making.  The authors have a particular interest in the relationship between scenarios and mainstream mathematical and statistical modelling.  It is tempting to put probabilities onto scenarios but there are at least two difficulties.  First, the scenarios may not comprehensively fill out the space of possibilities.  Second, probability is seen by some as an inappropriate tool in areas where probabilities become possibilities and the data aren't available to do detailed modelling. \par

Despite these problems, there remains some obvious distinctions.  For example, there are scenarios which cover events over which we may have no control and those which are firmly part of short or long term strategies. In the first case, we may only be left with mitigation and, in the second, where we have control, we still will have technical and technical, political and, today, ethical problems to overcome. It is surprising how many of the scenarios are common across different sectors.  This is because there are some that affect us all, particularly those related to climate change and allocation of resources in what may be a future with increased scarcity. \par

There is, as yet, no really coherent theory of scenarios, despite the fact that they are in increasing use.  One is reminded of a time when quality, reliability and risk were incorporated into management, with departments, champions, high level responsibility and training.  Risk, for example, is incorporated into corporate governance, and the black belts of quality improvements remain with us.  The use of scenarios reminds one of the early days of other soft system methodologies.  With risk, for example, we have fairly soft system management, but also heavy duty, model based methodologies in structural engineering and finance.  Today, we have the same issue with scenarios.  On the one hand, their use as a fairly soft management tool and, on the other, as we have suggested, their incorporation in stress testing mathematical models in decision support. Whether or not we are able to produce a coherent methodology, there is no doubt that companies, institutions and governments should make scenarios part of their organisational culture.  Just as we are familiar with the quality and safety cultures, we need to have a scenario culture, which is always by our side and affects our behaviour. Scenarios are a systems issue and, in developing and modelling the system, should be part of the structure.  Whether or not we put probabilities on them, they should form part of the foundation for our thinking.  A scenario culture will concentrate the mind and particularly the imagination on what might occur and what we do or do not have control over.  Like a safe driver, always attentive to what may happen as the road unfolds before them. \par 

\appendix
\section{Scenarios in practice}
The following sections describe the three sets of scenarios referred to in this paper.  

\subsection*{IPCC Representative Concentration Pathways}
The Intergovernmental Panel on Climate Change (IPCC) defines four scenarios on the basis of trajectories of radiative forcing in the Earth's atmosphere caused by emissions of greenhouse gases \cite{van2011representative}.  Scenarios were first produced on this basis in 2014 and replaced the scenarios defined in the Special Report on Emissions Scenarios first published in 2000. The scenarios are differentiated based on different assumed radiative forcing values in 2100.  In each scenario, a pathway defining level of radiative forcing in the years up to 2100 is defined.  Each scenario is summarised below.

\subsubsection*{RCP2.6}
RCP2.6 \cite{van2007stabilizing} is considered to be a best case scenario in terms of radiative forcing levels created by carbon emissions.  Under this scenario, radiative forcing levels  peak in 2040 at $3.1 W/m{2}$ and gradually fall to $2.6 W/m{2}$ by 2100.  To achieve this, very significant and ambitious reductions in carbon emissions would be required.

\subsubsection*{RCP4.5}
RCP4.5 is a scenario in which radiative forcing reaches a peak of $4.5 W/m{2}$ in 2100 and stabilises shortly after \cite{clarke2007scenarios}.  To reach this target, it is believed that significant switches in energy use to electricity would be required along with other lower emissions technologies and carbon capture \cite{thomson2011rcp4}.

\subsubsection*{RCP6}
RCP6 is a scenario in which growth in green technology and strategies enable the radiative forcing to stay below $6.0 W/m{2}$ until 2100 and stabilise shortly after \cite{hijioka2008global}.  The scenario is designed to represent a case in which levels of emissions are intermediate.  Over this period, there is still heavy reliance on fossil fuels, there is intermediate energy intensity and increasing use of croplands. Emissions peak in 2060 at 75 percent above today's level and reduce to 25 percent above by 2100.

\subsubsection*{RCP8.5}
RCP8.5 represents a scenario in which global greenhouse gas emissions continue to grow up to 2100 leading to extremely high radiative forcing \cite{riahi2007scenarios}. The scenario is designed to represent a case in which there are no policy changes aimed at reducing emissions.  Under the scenario, greenhouse gas emissions are three times that of today's with a high reliance on fossil fuels, population increase and a low rate of technology development.

\subsection*{National Grid Scenarios}
The UK National Grid defines energy scenarios on a yearly basis for use in government policy and for other energy planning \cite{NG_scenarios_2018}.  Previous to 2018, the scenarios were based on different levels of economic growth and government effort to decarbonise.  In 2018, a different approach was taken in which each scenario was defined with different assumptions about the level of decentralisation and the speed of decarbonisation.  Each of the four scenarios is summarised below:

\subsubsection*{B.1. Community Renewables}
Under this scenario, a relatively high level of decentralisation is assumed.  The UK government's commitment to reduce carbon emissions by eighty percent by 2050 is met.  There is extensive use of smart technology to manage peak loads and energy efficiency improves such that EU targets are met.  The government succeeds in its target to end all sales of petrol and diesel powered cars and electrical vehicles become the most popular form of private transportation.  For heavy goods vehicles, significant progress is made towards replacing natural gas with hydrogen.  Significant changes regarding the heating of homes have been made and heat pumps are the dominant technology.  There is also a significant role for district heating.  Electricity supply is dominated by solar and wind power, combined with efficient storage.  Gas is still important in the short to medium term but is eventually largely replaced with green gas. \par

\subsubsection*{B.2. Two Degrees}
The Two Degrees scenario meets the UK government's targets of reducing the carbon emissions by 80 percent by switching to larger and more centralised technologies.  Under this scenario, electricity demand is greatly reduced due to hydrogen heating.  Electrical appliances are generally much more energy efficient.  Electric vehicles become the most popular mode of private transport whilst demand for public transport also grows.  Hydrogen powered commercial vehicles, replacing natural gas powered vehicles, also become more widespread.   A large effort is made to improve the thermal efficiency of homes whilst the dominant heat sources are gas boilers, district heating and heat pumps.  Energy generation, primarily from wind and nuclear, is based on the transmission network itself, improving efficiency.  North Sea gas still plays an important role and some green gas is available. \par

\subsubsection*{B.3. Steady Progression}
Under this scenario, the government's 2050 decarbonisation target is not met but steady progress is made.  Improvements in the energy efficiency of appliances and the electrification of heat is slower than under the Community Renewables and Two Degrees scenarios but there is a large increase in the number of electric vehicles, giving an important role to smart technology in managing peak demand.  Gas powered vehicles continue to play an important role in the commercial sector.  Most homes still rely on the use of gas boilers and there is little improvement in energy efficiency or the use of heat pumps.  Electricity is generally generated on a large scale, rather than locally but there is development of nuclear power and offshore wind.  Gas still plays an important role with shale gas providing the majority of the supply. \par

\subsection*{Civil Litigation and Insurance}
Technological innovation in the automotive industry is concerning drivers, insurers and their respective legal representatives. At a general level, the rapid increase in electrical and/or driverless vehicles is the subject of much discussion.  At one end, it is an example of the risks on technological change.  Another such example is the possible psychological damage from social media and online gambling. \par

In collaboration with partners, the authors have been discussing the advent of driverless cars.  One issue, for example, is duty of care which is an ingredient in the definition of liability if there is an accident.  Who is liable, the driver or the designer of the vehicle?  Below, we list a number of scenarios, drawing on our own discussion and rather recent research in the area.

\subsubsection*{C.1.}
Uptake of driverless cars is low but this leads to an increase in the accident rate and a large number of expensive lawsuits concerning responsibility.  This leads to insurance premiums that are prohibitively expensive for many.  As a result, public opinion turns against driverless cars and a movement develops for more heavy regulation and/or a complete ban. \par

\subsubsection*{C.2}
Uptake of driverless cars is moderate and well regulated and the accident rate stays largely the same. A clear legal framework is set out regarding the apportioning of responsibility for accidents and insurance premiums are similar for both driverless and human-driven cars. \par

\subsubsection*{C.3}
Driverless cars are perceived as safe by the general public and public good arguments are used by regulators to allow their widespread introduction.  This leads to a sharp decrease in accidents and insurance premiums are cheaper as a result.  Human-driven cars are predicted by many to be obsolete within a decade. \par

\bibliographystyle{abbrv}
\bibliography{bibliography}

\end{document}